\UseRawInputEncoding
\documentclass[aps,prl,reprint]{revtex4-1}
\usepackage{amsthm,amsmath,amssymb}
\usepackage{graphicx}
\usepackage{epstopdf}
\usepackage{mathrsfs}
\usepackage{float}
\usepackage[colorlinks=true,citecolor=blue,dvipdfm]{hyperref}

\begin{document}


\title{Theory of Critical Phenomena with Memory}


\author{Shaolong Zeng, Sue ping Szeto, Fan Zhong}
\thanks{Corresponding author: stszf@mail.sysu.edu.cn}
\affiliation{State Key Laboratory of Optoelectronic Materials and Technologies, School of Physics, Sun Yat-Sen University, Guangzhou 510275, People's Republic of China}


\date{\today}

\begin{abstract}
Memory is a ubiquitous characteristic of complex systems and critical phenomena are one of the most intriguing phenomena in nature. Here, we propose an Ising model with memory and develop a corresponding theory of critical phenomena with memory for complex systems and discovered a series of surprising novel results. We show that a naive theory of a usual Hamiltonian with a direct inclusion of a power-law decaying long-range temporal interaction violates radically a hyperscaling law for all spatial dimensions even at and below the upper critical dimension. This entails both indispensable consideration of the Hamiltonian for dynamics, rather than the usual practice of just focusing on the corresponding dynamic Lagrangian alone, and transformations that result in a correct theory in which space and time are inextricably interwoven, leading to an effective spatial dimension that repairs the hyperscaling law. The theory gives rise to a set of novel mean-field critical exponents, which are different from the usual Landau ones, as well as new universality classes. These exponents are verified by numerical simulations of the Ising model with memory in two and three spatial dimensions.
\end{abstract}


\maketitle
Equilibrium critical phenomena are intriguing and has been well understood~\cite{Mask,Cardyb,Justin,amitb,Vasilev}. Classical critical dynamics is usually imposed by a Langevin equation with a Gaussian noise on the basis of instantaneous interactions and has also been well established~\cite{Hohenberg,Folk,Justin,Vasilev,Tauber}. Quantum critical dynamics of closed systems is determined by their Hamiltonians and can be obtained from analytical continuation through a quantum-classical correspondence~\cite{Sachdev}. However, open quantum systems may contain temporally nonlocal interactions when their coupled environmental degrees of freedom are integrated away~\cite{Sudip,Werner,Weiss,Yin3, Caruso,Breuer,Brown,Bulla,Winter,Kirchner,Sper,DeFili,Wang}. In fact, such a long-range temporal correlation or memory is ubiquitous in complex systems such as reaction-diffusion systems~\cite{Schulz,Tarasov}, disordered systems~\cite{Bouchaud,Metzler,West,Keim,Tsim,Maso,Maso1,Trimper02,Schutz,Trimper,Mok,Scalliet,Jack}, active matters~\cite{Nar,Loz}, epidemic spreading~\cite{Pastor,vanMie,Glee,Lin}, and many others~\cite{Zaslavsky,Rang,Beran,Wunsch,Valag,Valag1}, not to mention brain science and biology~\cite{Murray,Murray2,Freeman,Kappler,Zhang,Jiang} in general. It can result in effects such as anomalous diffusion that are qualitatively different from those exhibited by a memoryless system. What then will one expect memory or non-Markovian dynamics to bring to critical behavior?

To this end, we study a model with a time-dependent Hamiltonian,
\begin{equation}
\mathcal{H}_I(t)=-J\sum_{t_1<t}\sum_{<ij>}\frac{s_{i}(t)s_{j}(t_{1})}{(t-t_{1})^{1+\theta}},\label{ising}
\end{equation}
which extends the standard Ising model with a nearest-neighbor coupling constant $J$ to account for a long-range temporal interaction parameterizing by a constant $\theta>0$. In Eq.~\eqref{ising}, the Ising spin $s_i=\pm1$ at site $i$ may model, besides spins~\cite{Werner}, the phase of a Josephson junction in an array~\cite{Sudip}, a density of a chemical species in intracellular processes~\cite{Zhang} or complex reaction-diffusion systems~\cite{Schulz,Trimper}, positions of active particles~\cite{Nar,Loz}, or even a memorized pattern in a Hopfield neuron network~\cite{Jiang}, and can be generalized to more realistic forms. We have chosen a power-law decay as a simplification of various possible memory types~\cite{Sudip,Werner,Weiss,Yin3,Schulz,Tarasov,Valag,Valag1,Zaslavsky,Bouchaud,Metzler,West,Keim,Tsim,Maso,Maso1,Trimper02,Schutz,Trimper,Mok,Scalliet,Jack,Nar,Loz,Pastor,vanMie,Glee,Lin,Caruso,Breuer,Brown,Bulla,Winter,Kirchner,Sper,DeFili,Wang,Rang,Beran,Wunsch,Murray,Murray2,Freeman,Kappler,Zhang,Jiang}. The reason is that it is an analogy of a spatial counterpart~\cite{Fisher} and thus the single parameter $\theta$ anticipatorily enables us to classify different universality classes, which can then provide clues to characteristics of interactions in real systems. Such an interaction may be realized at least in artificial systems noticing that the decay exponent of the spatial counterpart can even be continuously tuned in trapped ions~\cite{Jurc,Britton,Islam1,Richerme,Bohnet,Yang}. To simulate the model, we employ Monte Carlo with Metropolis algorithm~\cite{MC,landaubinder}, which is interpreted as dynamics~\cite{Glauber,landaubinder} with the time unit of the usual Monte Carlo step per spin. However, a run of successive time steps of whatever length does not obviously converge to a time-dependent distribution $\exp\{-\mathcal{H}_I(t)/T\}$ due to the inherent memory in the model, where $T$ is the temperature in units of Boltzmann's constant. Rather, the ensemble average over different runs do because of the Markovian nature of the algorithm at every identical moment among different runs. In fact, such real ensemble averages have been successfully exploited to deal with Hamiltonian with a time-dependent term in finite-time scaling (FTS)~\cite{Gong,Zhong11,Feng}, in which scaling laws are found to be satisfied even if the fluctuation-dissipation theorem is not~\cite{Feng}. We will come back to FTS when we test the theory.

Theoretically, the long-wavelength low-frequency behavior of the model~\eqref{ising} is described by an effective na\"{\i}ve Hamiltonian $\mathscr{H}\equiv\int{dt}\mathcal{H}(t)$ with~\cite{noteh}
\begin{eqnarray}	
\mathcal{H}(t)&=&\int\!{d^dx}\left\{\frac{1}{2}\!\left[\tau\phi({\bf x},t)^2+(\nabla\phi({\bf x},t))^2\right]+\frac{1}{4!}u\phi({\bf x},t)^4 \right.\nonumber\\
&+&\left.\!\frac{1}{2}v\int_{-\infty}^{t}{dt_1}\frac{\phi({\bf x},t)\phi({\bf x},t_1)}{(t-t_1)^{1+\theta}}-h({\bf x},t)\phi({\bf x},t)\right\},\label{hscr}
\end{eqnarray}
which is the usual $\phi^4$ theory for an order-parameter field $\phi({\bf x},t)$ at position ${\bf x}$ and time $t$, where $\tau$, $v$, $u$, $h$, and $d$ denote a reduced temperature, the strength of the long-range interaction, a coupling constant, an additional ordering field, and the spatial dimensionality, respectively. In accordance with the Metropolis method, dynamics is governed by the Langevin equation,
\begin{equation}\label{2}
	\partial\phi({\bf x},t)/\partial t=-\lambda\delta\mathscr{H}/\delta \phi+\zeta
\end{equation}
with a kinetic coefficient $\lambda$ and a Gaussian white noise $\zeta$ of zero mean $\langle\zeta({\bf x},t)\rangle=0$ and correlator
\begin{equation}\label{3}
    \langle\zeta({\bf x},t)\zeta({\bf x}_1,t_1)\rangle=2\lambda\delta({\bf x}-{\bf x}_1)\delta(t-t_1)
\end{equation}
under the aforemention ensemble average. This dynamics has been well-known to be equivalent to a dynamic Lagrangian~\cite{Janssen79,Janssen,Folk,Justin,Vasilev,Tauber},
\begin{equation}\label{lcal}	
		\mathcal{L}=\!\int\!{dt d^dx}\left\{\tilde{\phi}\!\left[\partial\phi/\partial t+\lambda \delta\mathscr{H}/\delta\phi\right]\!-\!\lambda\tilde{\phi}^2\right\},
\end{equation}
where $\tilde{\phi}$ is an auxiliary response field~\cite{martin}.

The Hamiltonian Eq.~\eqref{hscr} is similar to its spatial counterpart, which correctly describes critical behavior of long-range spatial interaction systems. In particular, its critical behavior is divided into three regimes, a classical regime, a nonclassical regime, and the original short-ranged dominated regime~\cite{Fisher,Sak}. Here, surprisedly, we find that the na\"{\i}ve theory of its temporal counterpart is not correct any more. We show below that such a theory violates a hyperscaling law~\cite{Mask,Cardyb,Justin,amitb,Vasilev,scalinglaw}
\begin{equation}
\beta/\nu+\beta\delta/\nu=d,\label{hyper}
\end{equation}
for \emph{all} $d$, once the memory dominates, where $\beta$, $\nu$, and $\delta$ are critical exponents. This is in sharp contrast to the usual Landau mean-field exponents which break Eq.~\eqref{hyper} only above the upper critical dimension $d_c$. Since this scaling law ensures a fundamental thermodynamic relation $M=-(\partial f/\partial h)_{\tau}$ between an order parameter $M$ and $h$ through the singular part of a time-dependent free energy,
\begin{equation}
f(\tau,h,t)=b^{-d}f(\tau b^{1/\nu}, hb^{\beta\delta/\nu},tb^{-z}),\label{ftht}
\end{equation}
arising from $\exp\{-\mathcal{H}_I(t)/T\}$, violation of the hyperscaling law then implies violation of the thermodynamics~\cite{notef}, where $z$ is the dynamic critical exponent. Moreover, to correctly describe the critical behavior of the model~\eqref{ising}, we demonstrate that a proper dimensional correction of the theory is indispensable. This leads to a series of novel results in the corrected theory such as the role of $\mathcal{H}$ in dynamics, the intimate relation between space and time, and a set of novel mean-field critical exponents different from the usual Landau ones.

We will utilize the renormalization-group (RG) technique. To this end, we write Eqs.~\eqref{lcal} and~\eqref{hscr} together as
\begin{eqnarray}\label{lcalk}	
		\mathcal{L}=\!\int\!\!{dt d^dx}\left\{\tilde{\phi}\!\left[\lambda_1\frac{\partial\phi}{\partial t}+\tau\phi-K\nabla^2\phi+\frac{1}{3!}u\phi^3 \right.\right.\nonumber\\
\left.\left.+v\int{dt_1}\frac{\phi(t_1)}{(t-t_1)^{1+\theta}}-h-\lambda_2\tilde{\phi}\right]\right\},
\end{eqnarray}
where we have replaced $\lambda$ by three constants $\lambda_1$, $\lambda_2$, and $K$ to account for possible different renormalizations of the terms. The RG method is first to integrate out the degrees of freedom within the momentum shell $\Lambda/b\le k\le\Lambda$ and then to rescale $k$ to $k_r=kb$ so that the cutoff $\Lambda$ of the remaining degrees of freedom recovers its original value, where $b>1$ is a rescaling factor and $r$ denotes renormalized quantity. To keep the form of $\mathcal{L}$ after renormalization, we also rescale
\begin{equation}
x_r=xb^{-1},~t_r=tb^{\gamma_t},~\phi_r=\phi b^{\gamma_{\phi}},~\tilde{\phi}_r=\tilde{\phi}b^{\gamma_{\tilde{\phi}}},~h_r=hb^{\gamma_h},\label{xtrg}
\end{equation}
where the constant $\gamma_o$ is referred to hereafter as the dimension of $o$ defined as $o_r=ob^{\gamma_o}$. Critical exponents can be obtained through
\begin{equation}
\beta/\nu=\gamma_{\phi},\qquad\beta\delta/\nu=\gamma_h,\qquad z=-\gamma_t,\label{bnbdn}
\end{equation}
via scaling hypothesis~\cite{Mask,Cardyb,Justin,amitb,Vasilev} similar to Eq.~\eqref{ftht}. For an infinitesimal transformation, $b=1+dl$ with $dl\ll1$, keeping only terms to one-loop order~\cite{Mask,Cardyb,Justin,amitb,Vasilev}, we obtain
\begin{subequations}\label{rgdl}
\begin{eqnarray}
     \partial_lv&=&\left(d-\gamma_t-\gamma_{\phi}-\gamma_{\tilde{\phi}}+\theta\gamma_t\right)v,\label{rgdlv}\\     \partial_lu&=&(d-\gamma_t-3\gamma_{\phi}-\gamma_{\tilde{\phi}})u-3u^2I_2/dl,\label{rgdlu}\\
     \partial_l\ln K&=&d-2-\gamma_t-\gamma_{\phi}-\gamma_{\tilde{\phi}},\label{rgdlk}\\
     \partial_l\ln\lambda_{2}&=&d-\gamma_t-2\gamma_{\tilde{\phi}},\label{rgdll2}\\
     \partial_l\ln\lambda_{1}&=&d-\gamma_{\phi}-\gamma_{\tilde{\phi}}=\gamma_{\lambda_1},\label{rgdll1}\\
     \partial_l\ln h&=&d-\gamma_t-\gamma_{\tilde{\phi}}-\gamma_h,\label{rgdlh}\\
     \partial_l\ln\tau&=&d-\gamma_t-\gamma_{\phi}-\gamma_{\tilde{\phi}}+uI_1/2\tau dl,\label{rgdlt}
\end{eqnarray}
\end{subequations}
where $\partial_l o\equiv\partial o/\partial l=(o_r-o)/dl$ and the two integrals are defined as
\begin{eqnarray}\label{I1234}
I_1&=&\frac{1}{(2\pi)^{d+1}}\!\int_{\Lambda/b}^{\Lambda}{dk}\!\int{d\omega}C_0(k,\omega)\equiv\!\int_{k}^{b}\!\int_{\omega}\!C_0(k,\omega),\nonumber\\
I_2&=&\!\int_{k}^{b}\!\int_{\omega}\!C_0(k,\omega)G_0(-k,-\omega),
\end{eqnarray}
with the response $\langle\phi'({\bf k},\omega)\tilde{\phi}'(-{\bf k},-\omega)\rangle$ and correlation $\langle\phi'({\bf k},\omega)\phi'(-{\bf k},-\omega)\rangle$  functions~\cite{Janssen79,Janssen,Folk,Justin,Vasilev,Tauber,Zhongfp} being given by
\begin{equation}\label{g0}
G_0(k,\omega)^{-1}=\tau+Kk^2+v\Gamma(-\theta)(-i\omega)^{\theta}-i\lambda_1\omega
\end{equation}
($\Gamma$ is the gamma function) and $C({\bf k},\omega)=2\lambda_2G_0(k,\omega)G_0(k,-\omega)\equiv \lambda_2C_0(k,\omega)$, respectively. 
Since even the mean-field results of the na\^{\i}ve theory breach the fundamental law, we need essentially a Gaussian analysis without the results of the two integrals,.

We now analyse the fixed points of the RG equations and their properties. This is obtained by equating all equations in Eq.~\eqref{rgdl} with zero~\cite{noterg0}. There are two different cases. The first case is $v^*=0$ from Eq.~\eqref{rgdlv}, where the star denotes the fixed-point value. The memory is irrelevant and we recover the usual short-range interaction $\phi^4$ theory. Indeed, solving the fixed-point equations from Eqs.~\eqref{rgdlk}--\eqref{rgdlh} yields
\begin{equation}
\gamma_t^*=-2,\quad\gamma_{\phi}^*=(d-2)/2,\quad\gamma_{\tilde{\phi}}^*=\gamma_h^*=(d+2)/2,\label{sov0}
\end{equation}
and hence $u^*(4-d-3I_2u^*/dl)=0$ from Eq.~\eqref{rgdlu}. Standard textbook analysis~\cite{Mask,Cardyb,Justin,amitb,Vasilev} shows that the Gaussian fixed point $u^*=0$ is stable above $d_{c0}=4$ with the Gaussian dimensions, Eq.~\eqref{sov0}, and hence the correct Gaussian critical exponents $\beta/\nu$, $\beta\delta/\nu$, and $z$ from Eq.~\eqref{bnbdn}. The hyperscaling law~\eqref{hyper} is satisfied. However, the Landau mean-field exponents, $\beta/\nu=1$ and $\beta\delta/\nu=3$, the Gaussian exponents at $d_{c0}$, only fulfil Eq.~\eqref{hyper} at $d_{c0}$. For $d<d_{c0}$, since $I_2>0$, there exists a new fixed point $u^*=dl(4-d)/3I_2$ that controls the nonclassical regime in which the critical exponents in Eq.~\eqref{sov0} have corrections depending on $4-d$~\cite{Mask,Cardyb,Justin,amitb,Vasilev}.

For the long-range fixed point, $v^*\neq0$, one finds instead
\begin{equation}
u^*(6-2/\theta-d-3I_2u^*/dl)=0,\label{fpu}
\end{equation}
using the fixed-point equations from Eqs.~\eqref{rgdlv}--\eqref{rgdll2}. The same analysis shows that the Gaussian fixed point $u^*=0$ is now stable above
$d_{c}=6-2/\theta$
with the Gaussian dimensions now changing to
\begin{equation}\label{dimsol}
\gamma_t^*=-\frac{2}{\theta},\quad\gamma_{\phi}^*=\frac{d}{2}+\frac{1}{\theta}-2,\quad\gamma_{\tilde{\phi}}^*=\gamma_h^*=\frac{d}{2}+\frac{1}{\theta}.
\end{equation}
As a result, we have now $z=2/\theta$ and
\begin{equation}
\beta/\nu+\beta\delta/\nu=d-2+2/\theta.\label{shadow}
\end{equation}
Equation~\eqref{shadow} indicates that the hyperscaling law is violated radically for all $d$ except $\theta=1$ and therefore the na\"{\i}ve theory cannot be right.

We note that both sets of the exponents, Eqs.~\eqref{sov0} and~\eqref{dimsol}, give rise to $\gamma_{\tau}^*=d-\gamma_t^*-\gamma_{\phi}^*-\gamma_{\tilde{\phi}}^*=2$ in Eq.~\eqref{rgdlt}. This indicates that $\tau$ increases with $b$ and thus is relevant. Moreover, the Gaussian exponent $\nu=1/\gamma_{\tau}^*=1/2$ for both the short-rang and the long-range fixed points~\cite{Mask,Cardyb,Justin,amitb,Vasilev}. Similarly, $\gamma_{\lambda_1}^*=2-2/\theta$ for the long-range fixed point from Eq.~\eqref{rgdll1}. Accordingly, it is negative for $\theta<1$ and therefore is irrelevant.

We now focus on the theory with memory and show that the violation of the hyperscaling law originates from the dimension of $\mathcal{H}$. Indeed, using Eq.~\eqref{xtrg} and the short-range results, Eq.~\eqref{sov0}, one sees that all terms in $\mathcal{H}$ have a zero dimension except inconsistently the memory term. However, with the long-range results, Eq.~\eqref{dimsol}, all terms then share an identical dimension
\begin{equation}\label{dscr}
    \gamma_{\mathcal{H}}=2-2/\theta=\gamma_a,
\end{equation}
which is minus the extra term in Eq.~\eqref{shadow}. This prompts us to eliminate $\gamma_{\mathcal{H}}$ in order to save the hyperscaling law and the theory.

To this end, we multiply $\mathscr{H}$ by a constant $a$ such that $\gamma_{\mathcal{H}'}=-\gamma_a+\gamma_{\mathcal{H}}=0$. However, introducing an additional scaling field to the theory breaks the correspondence between Eqs.~\eqref{ising} and~\eqref{hscr}. This means that $a$ just contributes a factor $b^{-\gamma_a}$ (note the special minus sign to emphasize its speciality) instead of changing to $a_r$ upon rescaling. $\zeta$ ought to be multiplied by $a$ accordingly. Its correlator in Eq.~\eqref{3} is then proportional to $a^2$. Consequently, $\mathcal{L}$, Eq.~\eqref{lcal}, becomes
\begin{equation}\label{10}
\mathcal{L}'=\int{dtd^dx}\left[\tilde{\phi}({\bf x},t)a\delta\mathscr{H}/\delta\phi-a^2\tilde{\phi}({\bf x},t)^2\right],
\end{equation}
where we have omitted $\lambda$ and the irrelevant time derivative term for $\theta<1$. One sees from Eq.~\eqref{10} that if one defines $\tilde{\phi}_1=a\tilde{\phi}$, $\mathcal{L}$ recovers its original form. This is reasonable as $\tilde{\phi}$ originates from integration out of $\zeta$~\cite{martin}. The replacement implies that one ought to replace $\gamma_{\tilde{\phi}}$ with $\gamma_{\tilde{\phi}}+\gamma_a$ in Eq.~\eqref{rgdl} and thus the new hyperscaling law~\eqref{shadow} does hold.

However, this is not enough. To see why, we make a transformation,
\begin{equation}
\phi=\phi' a^{-n},\qquad\tilde{\phi}=\tilde{\phi}' a^{-\tilde{n}},\label{phipi}
\end{equation}
with two constants $n$ and $\tilde{n}$ and obtain
\begin{eqnarray}\label{hlpi}
    \mathcal{H}'&=&a^{1-2n}\int_k\left[\frac{1}{2}G_0^{-1}|\phi'|^2+\frac{1}{4!}a^{-2n}u\mathcal{H}_u-a^{n}h\phi'\right],\nonumber\\
    \mathcal{L}'&=&\int_{\omega,k}\tilde{\phi'}\left[ a^{1-n-\tilde{n}}G_0^{-1}\phi'+\frac{1}{3!}a^{1-3n-\tilde{n}}u\frac{\delta\mathcal{H}_u}{\delta\phi'}\right.\nonumber\\
&\quad&\qquad\qquad\qquad\qquad\quad\;\:\:-\:a^{1-\tilde{n}}h-a^{2-2\tilde{n}}\tilde{\phi'}\bigg],
\end{eqnarray}
after Fourier transform, where $\mathcal{H}_u$ denotes the Fourier transform of $\int\!dtd^dx\phi^4$ with one frequency and wavenumber integration extracted, and we have used identical symbols for both direct spacetime and its reciprocal and suppressed the arguments for clarity.

In Eq.~\eqref{hlpi}, we have drawn $a^{1-2n}$ out of the integration for $\mathcal{H}'$. In this way, we find the perturbation contribution to the renormalized vertex, $\tilde{u}$, becomes
\begin{equation}\label{upih}
      \tilde{u}=u\left\{1-\frac{3}{4}a^{1-4n}u\!\int_k\!G_0(k,0)^2\right\},
\end{equation}
to one-loop order for all $t$, $t_1$, and $(t-t_1)$ in Eq.~\eqref{hscr} approaching infinity. It can be shown combinatorially that all higher orders contain powers of $a^{1-4n}u$~\cite{nh}. Therefore, if we choose $n=1/4$, $a$ does not contribute to the renormalization. This is also true for other parameters such as $\tau$ and substantially disentangles the involvement of $a$ in the theory.

For $\mathcal{L}'$ in Eq.~\eqref{hlpi}, the transformation Eq.~\eqref{phipi} represents only a redistribution of $\gamma_a$ between the fields and must not introduce an overall factor. Accordingly, from Eq.~\eqref{hlpi}, we have $n=1-\tilde{n}$, or $\tilde{n}=3/4$ to ensure identical $h$ transformations ($h'=ha^{1/4}$), propagators, and $u$ for $\mathcal{H}'$ and its $\mathcal{L}'$. Although the last term in $\mathcal{L}'$ leads to an $a$-dependent correlation function $C({\bf k},\omega)=a^{2n}C_0(k,\omega)$, one can again show that all $a$ are exactly cancelled in the two integrals in Eq.~\eqref{I1234} for the renormalized $\tau_r$ and $u_r$ to one-loop order and to all orders combinatorially~\cite{nl}. Conversely, without the transformation~\eqref{phipi}, $a$ could not serve as a simple constant only, since it would be differently involved in $\mathcal{H}$ and $\mathcal{L}$, which are then inconsistent with each other.

We note that for $\mathcal{H}'$ and $\mathcal{L}'$ to be really consistent, each integration over wavenumber for $\mathcal{L}'$ and hence in Eq.~\eqref{I1234} must also contain the same $a^{1-2n}=a^{1/2}$ factor. However, it must be exactly cancelled by $a^{-1/2}$ for the frequency integration. The same happens also to the integrations in $\mathcal{H}_u$. Accordingly, the correct theory in place of Eqs.~\eqref{hscr} and~\eqref{lcal} is
\begin{eqnarray}
\mathcal{L}'&=&\int{\!\left(dta^{-\frac{1}{2}}\right)\!\left(d^dxa^{\frac{1}{2}}\right)}\tilde{\phi'} \left\{\tau\phi'-\nabla^2\phi'+\frac{1}{3!}a^{-\frac{1}{2}}u\phi'^3\right.\nonumber\\
&\quad&\qquad\left.+v\int_{-\infty}^{t}{dt_1}\frac{\phi'({\bf x},t_1)}{(t-t_1)^{1+\theta}}-h'-a^{\frac{1}{2}}\tilde{\phi'}\right\}.\label{lpic}
\end{eqnarray}

Equation~\eqref{lpic} and its associated $\mathcal{H}'$ constitute our corrected theory of critical phenomena with memory. The RG transformations of the theory are similar to Eq.~\eqref{rgdl} except that the transformations for $u$ and $\lambda_2$ need add and subtract $\gamma_a/2$, respectively, because of the extra factors. The fixed-point solution for $u$ is again Eq.~\eqref{fpu} while
\begin{equation}\label{phihp}
\begin{split}
  \gamma_t^*=-2/\theta,\quad\gamma_{\phi'}^*=d/2+1/2\theta-3/2,\qquad\\
  \gamma_{\tilde{\phi}}^*=d/2+3/2\theta-1/2\quad\gamma_{h'}^*=d/2+1/2\theta+1/2,
\end{split}
\end{equation}
which leads to
\begin{equation}
\beta/\nu+\beta\delta/\nu=d-\gamma_a/2=d_{\rm eff},\label{shadowpi}
\end{equation}
using Eq.~\eqref{bnbdn}. Although Eq.~\eqref{shadowpi} appears to violate both Eqs.~\eqref{hyper} and~\eqref{shadow}, $d_{\rm eff}$ is just the dimension of $d^dxa^{1/2}$ and hence an effective spatial dimension. Accordingly, the hyperscaling law~\eqref{hyper} holds again in it! This is corroborated by the fact that the coupling term $a^{-1/2}u$ now scales as $b^{d_c-\gamma_a/2-d}\equiv b^{d_{c{\rm eff}}-d}$. In other words, the effective upper critical dimension $d_{c{\rm eff}}$ changes consistently to $d_c-\gamma_a/2=5-1/\theta$. Similarly, the usual hyperscaling law ought to be modified to
$d_{\rm eff}\nu=(d-\gamma_a/2)\nu=2-\alpha$ accordingly, while the others such as $\gamma=(2-\eta)\nu$~\cite{Mask,Cardyb,Justin,amitb,Vasilev} remain unchanged, where $\alpha$ and $\gamma$ are the critical exponents for specific heat and susceptibility $\chi$, respectively. From Eq.~\eqref{dscr},  $d_{\rm eff}$ is always bigger than $d$ in the memory-dominant regime, a fact which is qualitatively reasonable as the temporal correlation inevitably suppresses spatial fluctuations and contributes an effective spatial dimension.

Moreover, in the memory-dominated regime, the correct mean-field critical exponents, Eq.~\eqref{phihp} at $d_c$, become
\begin{equation}
\beta/\nu=3/2-1/{2\theta},\qquad \beta\delta/\nu=7/2-1/2\theta,\label{bnd}
\end{equation}
though $\nu=1/2$ since $\gamma_{\tau}^*=2$ too and $\gamma=1$ as usual. For $\theta=1/2$ and $2/3$, for example, $d_c=2$ and $3$ and $\beta/\nu=1/2$ and $3/4$, respectively, different from the Landau result $\beta/\nu=1$ obtained from both Eq.~\eqref{sov0} and Eq.~\eqref{dimsol}.

In addition, since the fixed-point equation~\eqref{fpu} remains intact, $d_c$ again separates the Gaussian regime above it from and nonclassical regimes below it. Also, from Eq.~\eqref{phihp}, $\gamma_{\lambda_1}^*=2-2/\theta$ too, the division of the short-range and long-range regimes remains intact. In fact, to $\epsilon^2=(d_c-d)^2$, we find that this division shifts to $\theta=1-\kappa$ with $\kappa=\ln(4/3)\epsilon^2/18$~\cite{Zeng}, similar to the spatial counterpart~\cite{Sak}.

Turning to $a^{-1/2}$ in the temporal integration in Eq.~\eqref{lpic}, we need to transform $t$ to $t'=a^{-1/2}t$, which just accounts for the last extra $a^{1/2}$ in Eq.~\eqref{lpic} because $a^{1/2}\delta(t)=\delta(t')$ in Eq.~\eqref{3}. Indeed, an RG analysis with $t'$ instead of $t$ results again in Eq.~\eqref{phihp} with a new
$z'=-\gamma_{t'}^*=z+\gamma_a/2$~\cite{ntc},
which indicates that part of the temporal dimension has been transferred to the space and which leads correctly to $G_0=b^2/[\tau_r+K_Rk^2_r+v_ra^{-\theta/2}\Gamma(-\theta)(-i\omega'_r)^{\theta}]$ upon neglecting the irrelevant $\lambda_1$ term in Eq.~\eqref{g0}, where $\omega'_r=a^{1/2}\omega b^{-\gamma_{t'}^*}$. A similar scaling obviously holds for $G_0$ in terms of $t$ as well.

We have seen that $\mathcal{H}$ plays an essential role due to Eq.~\eqref{dscr}. This draws attention to the eluded role played by $\mathcal{H}$ in dynamics, as nowadays one usually focuses solely on $\mathcal{L}$ for critical dynamics though $\mathcal{H}$ controls the equilibrium properties~\cite{Janssen79,Janssen,Folk,Justin,Vasilev,Tauber}. Since $a$ arises from the memory, the scaling laws thus connect both static and dynamic properties, demonstrating the intimate relation between space and time similar to quantum phase transitions but adjustable through $\theta$ in the memory-dominated regime.

To verify the exponents, we utilize FTS~\cite{Gong,Zhong11,Feng}. FTS is to change a controlling parameter in Hamiltonian linearly with time at a rate $R$ through a critical point and is perfect for dealing with the time-dependent Hamiltonian~\eqref{ising}. The driving imposes on the system a controllable external finite timescale that plays the role of system size in finite-size scaling and leads to FTS~\cite{Gong}. FTS has been successfully applied to many systems to efficiently study their equilibrium and nonequilibrium critical properties~\cite{Gong,Zhong11,Feng,Yin,Yin3,Liu,Huang,Liupre,Liuprl,Huang1,Pelissetto,Xu,Xue,Cao,Gerster,Li,Mathey,Yuan,Yuan1,Yuan2,Zuo,Clark,Keesling}.
Here, we change $T$ through $\tau=T-T_{c}=Rt'$~\cite{notetau} and choose $R$ as a variable in place of $t'$. Accordingly, Eq.~\eqref{ftht} becomes~\cite{Gong,Zhong11,Feng}
\begin{equation}\label{fthtl}
	f(\tau,h,R)=b^{-d_{{\rm eff}}}f(\tau b^{1/\nu},h'b^{\beta\delta/\nu},Rb^{r}),
\end{equation}
where $r$ is a rate exponent~\cite{Zhong02,Zhong06}. Since $t'$ scales as $t'b^{-z'}$, one finds
$r=z'+1/\nu=3+1/\theta$, 
which enables us to verify $z'$. Consequently, one arrives at the FTS forms
\begin{equation}\label{Mchi}
  M=R^{\beta/r\nu}f_{1}(\tau R^{-1/r\nu}),~
  \chi=R^{-\gamma/r\nu}f_{2}(\tau R^{-1/r\nu}),
\end{equation}
for $M$ and $\chi$ at $h'=0$, where $f_1$ and $f_{2}$ are scaling functions.
At the peak of $\chi$, $T_p=T_c+cR^{1/r\nu}$ with a constant $c$ satisfying $df_{2}(x)/dx|_{x=c}=0$ from Eq.~\eqref{Mchi}~\cite{Gong,Zhong11}. This yields $1/r\nu$ and $T_c$, at which $M_c\propto R^{\beta/r\nu}$ and $\chi_c\propto R^{-\gamma/r\nu}$ and thereby $\beta/r\nu$ and $\gamma/r\nu$. However, $T_c$ so estimated has a relatively large error and hence affects others. So, we apply the theoretical $1/r\nu$ to fit $T_c$ and then use it to determine other exponents and check consistently by curve collapsing.

\begin{figure}
\includegraphics[width=1\columnwidth]{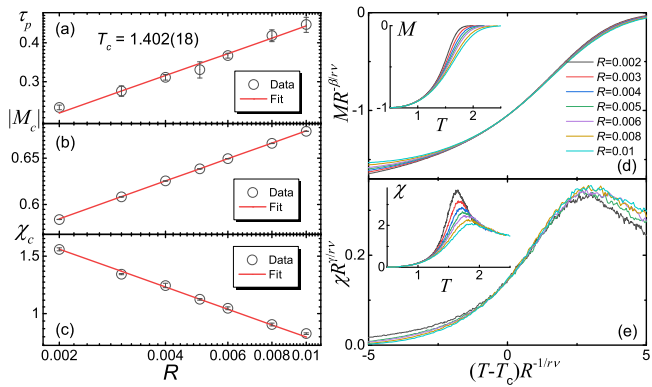}
\caption{(Color online) \label{fts2d05} (a) $\tau_p=T_p-T_c$, (b) $M_c$, (c) $\chi_c$, (d) and (e), rescaled $M$ and $\chi$, respectively, of the Ising model with memory with $\theta=0.5$ and ${\cal T}=20$ and subject to heating at a series of $R$ given in the legend for $40~000$ samples on $100\times100$ square lattices. The insets in (d) and (e) are the respective original curves. Each column shares identical abscissa axes. Double logarithmic scales are used in (a)--(c).}
\end{figure}
We simulate the model~\eqref{ising} with $J=0.2$ and a cutoff ${\cal T}$ of the long-range interaction. Periodic boundary conditions are applied throughout and initial ordered states are carefully chosen to be far away from $T_c$ and thus do not affect results. In Fig.~\ref{fts2d05}, we display the results for the 2D model. We estimate, with the method mentioned, $T_{c}=1.402(18)$, which leads to Fig.~\ref{fts2d05}(a)--\ref{fts2d05}(c), where the straight lines have slopes of $1/r\nu=0.405(22)$, $\beta/r\nu=0.0941(4)$, and $-\gamma/r\nu=-0.372(9)$, close to the theoretical values of $2/5$, $1/10$, and $-2/5$, respectively. Note that if $t$ were not transformed to $t'$, the three exponents would be $1/3$, $1/12$, and $-1/3$, respectively, and further, if $\beta$ were $1/2$, $\beta/r\nu=1/6$, all far away from the numerical values. Moreover, these estimated exponents result in the rather good curve collapses in Figs.~\ref{fts2d05}(d) and~\ref{fts2d05}(e) around $T_c$, though far away from it, the collapses are not quite good near the peaks where fluctuations are large as expected~\cite{Yuan,Yuan2}. Similar results are obtained for ${\cal T}=15$~\cite{Zeng}. This is reasonable once the driven time is shorter than the cutoff memory time so that the latter becomes sub-leading. Similar condition applies to the lattice sizes as well~\cite{Zhong11,Feng,Huang}. Although these sub-leading factors may still contribute to the small difference of the numerical results, any logarithmic correction is found to deteriorate the collapses~\cite{Luijten97}.

\begin{figure}
\includegraphics[width=1\columnwidth]{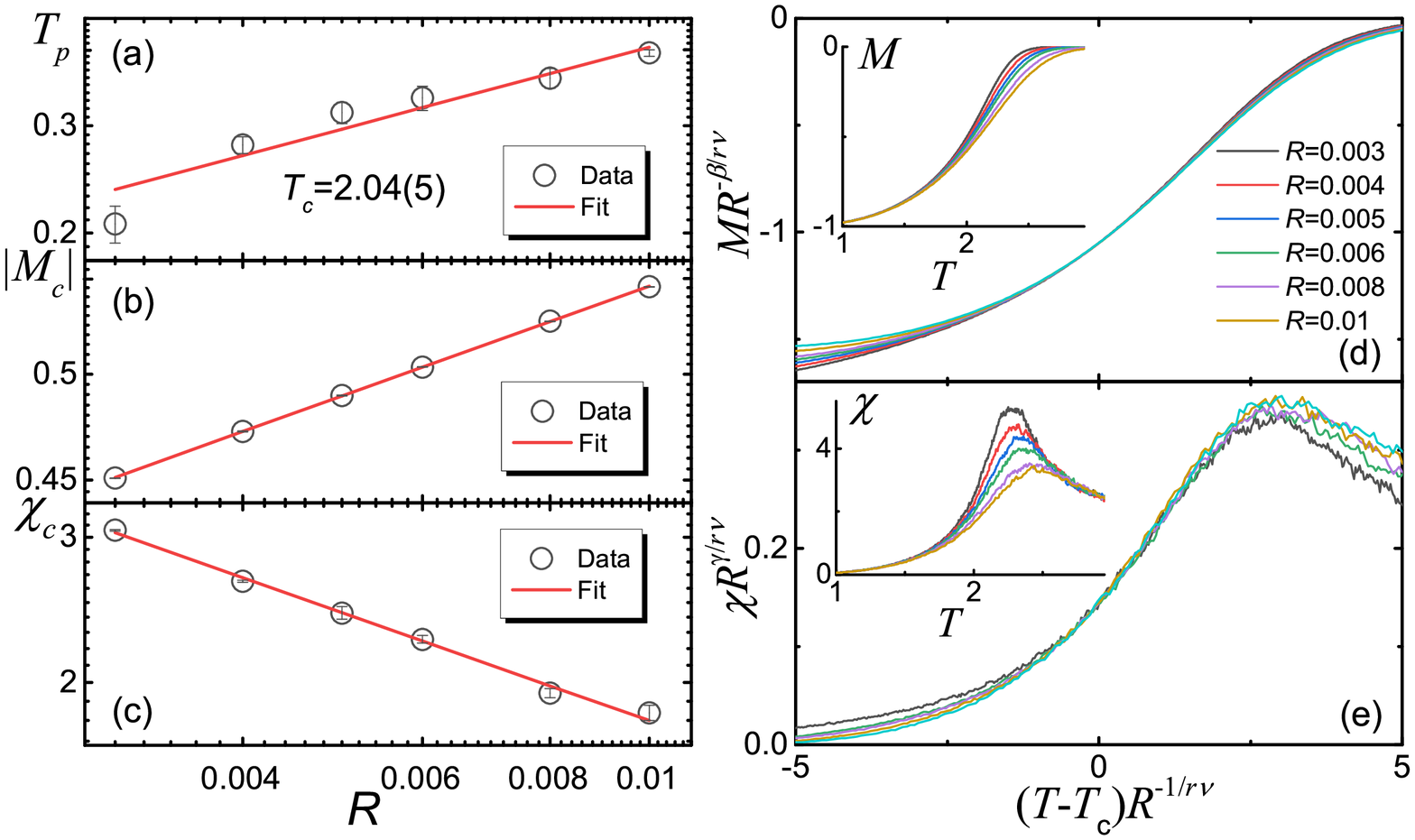}
\caption{(Color online) \label{fts3d05} Results of the 3D Ising model with $\theta=2/3$ and ${\cal T}=18$ for $40~000$ samples on $25\times25\times25$ simple cubic lattices, corresponding to those in Fig.~\ref{fts2d05}.}
\end{figure}
The 3D exponents are estimated to be $1/r\nu=0.45(6)$, $\beta/r\nu=0.1580(7)$, and $-\gamma/r\nu=-0.435(14)$ shown in Fig.~\ref{fts3d05}(a)--\ref{fts3d05}(c), again close to the theoretical values of $4/9$, $1/6$, and $-4/9$, respectively. They again yield rather good curve collapses in Figs.~\ref{fts3d05}(d) and~\ref{fts3d05}(e) around $T_c$.

Concluding, we have proposed a model and derived the corresponding theory for critical phenomena with memory arising from a temporal power-law decaying interaction. The theory contains a unique dimensional constant that originates from the dimension of the Hamiltonian, that inextricably interweaves space and time into an effective dimension, that serves to repair the radically violated hyperscaling law, that transforms the time, that changes the Gaussian and mean-field critical exponents such that the latter are different from the usual Landau ones, that draws attention to the eluded role played by the Hamiltonian in dynamics, that dramatically distinguishes the theory from its spatial counterpart, and yet, that is not a scaling field which alters the fixed points. New universality classes emerge.

\begin{acknowledgments}
This work was supported by the National Natural Science Foundation of China (Grant Nos. 11575297 and 12175316).
\end{acknowledgments}


\begin{thebibliography}{99}
\bibitem{scalinglaw} It is derived from the standard scaling laws $d\nu=2-\alpha$, $\alpha+2\beta+\gamma=2$, and $\gamma=\beta(\delta-1)$~\cite{Mask,Cardyb,Justin,Vasilev,amitb}.
\bibitem{notef} From Eq.~\eqref{ftht}, $\partial f/\partial h=b^{-d+\beta\delta/\nu}f'(\tau b^{1/\nu}, hb^{\beta\delta/\nu},tb^{-z})$, where $f'$ denotes the derivative of $f$ with respect to its second argument. Since $M$ scales as $Mb^{\beta/\nu}$, if the hyperscaling law~\eqref{hyper} holds, the thermodynamic relation $M=-\partial f/\partial h$ holds too. Otherwise, the thermodynamic relation cannot be valid.
\bibitem{noterg0} In the long-range fixed point $v^*\neq0$ and to the present one-loop order, Eq.~\eqref{rgdll1} is equal to $\gamma_{\lambda_1}=2-2/\theta$ rather than zero because it is irrlevant (see the text below). However, to higher orders, it is again equal to zero and leads to crossover, see~\cite{Zeng} for details.
\bibitem{nl} A one-particle irreducible graph with $m$ vertices has $(4m-4)/2=2m-2$ internal lines, among which there are $m-1$ $G_0$ lines and $m-1$ $C$ lines. As a result, $(ua^{-2n})^ma^{2n(m-1)}=u^{m-1}(ua^{-2n})$ for the four-leg $u$. For the two-leg $\tau$, the number of internal lines is $(4m-2)/2=2m-1$ and that of the $C$ lines is $2m-1-m+1=m$ because of the same number of $G_0$ lines. Therefore, $(ua^{-2n})^ma^{2nm}=u^{m}$.
\bibitem{ntc} Note that although $\gamma_t$ becomes $\gamma_{t'}=\gamma_t-\gamma_a/2$, $d$ also changes to $d-\gamma_a/2$. Consequently, $\gamma_h$ keeps unchanged.
\bibitem{notetau} For simplicity, we do not differentiate $\tau$ used hereafter and that before. Here $T_c$ is the real critical temperature that includes the contribution both from fluctuations and from the long-range temporal interaction, whereas the previous $\tau$ is well known to be defined as the distance to the mean-field critical temperture only.
\end{thebibliography}
\end{document}